\documentclass[journal,twoside,web]{IEEEtran}
\usepackage{amsmath,amsfonts}
\usepackage{algorithm}
\usepackage{array}
\usepackage[caption=false,font=normalsize,labelfont=sf,textfont=sf]{subfig}
\usepackage{textcomp}
\usepackage{stfloats}
\usepackage{url}
\usepackage{verbatim}
\usepackage{graphicx}
\usepackage{cite}
\usepackage[hidelinks]{hyperref}
\usepackage{url}
\usepackage{pifont}
\usepackage{cuted}
\usepackage{multirow}
\usepackage{booktabs}
\usepackage[normalem]{ulem}
\usepackage{booktabs}

\def\ii{{\hat{\imath}}}												
\def\ij{{\hat{\jmath}}}												
\def\ik{{\hat{\kappa}}}												
\def\bH{\mathbb{H}}														
\def\bR{\mathbb{R}}														

\title{Generalizing Medical Image Representations \\via Quaternion Wavelet Networks}
\author{Luigi~Sigillo,~\IEEEmembership{Graduate Student~Member,~IEEE}, Eleonora~Grassucci, \\ Aurelio~Uncini,~\IEEEmembership{Senior~Member,~IEEE}, and~Danilo~Comminiello,~\IEEEmembership{Senior Member,~IEEE} \thanks{Authors are with the Department of Information Engineering, Electronics and Telecommunications (DIET), Sapienza University of Rome, Italy. This work was partly supported by “Ricerca e innovazione nel Lazio - incentivi per i dottorati di innovazione per le imprese e per la PA - L.R. 13/2008” of Regione Lazio, Project “Deep Learning Generativo nel Dominio Ipercomplesso per Applicazioni di Intelligenza Artificiale ad Alta Efficienza Energetica”, under grant number 21027NP000000136, and by the European Union under the National Plan for Complementary Investments to the Italian National Recovery and Resilience Plan (NRRP) of NextGenerationEU,  Project PNC 0000001 D3 4 Health, (Digital Driven Diagnostics, prognostics and therapeutics for sustainable Health care) - SPOKE 1 Clinical use cases and new models of care supported by AI/E-Health based solutions - CUP B53C22006120001. The work of A. Uncini has been partly supported by the European Union under the NRRP of NextGenerationEU, partnership on “Future Artificial Intelligence Research” (PE00000013 – SPOKE 5 - CUP B53C22003980006 - FAIR: High Quality AI).
}

\thanks{Corresponding author's email: luigi.sigillo@uniroma1.it.}}

\begin{document}

\maketitle

\begin{abstract}
Neural network generalizability is becoming a broad research field due to the increasing availability of datasets from different sources and for various tasks. This issue is even wider when processing medical data, where a lack of methodological standards causes large variations being provided by different imaging centers or acquired with various devices and cofactors.
To overcome these limitations, we introduce a novel, generalizable, data- and task-agnostic framework able to extract salient features from medical images. The proposed quaternion wavelet network (QUAVE) can be easily integrated with any pre-existing medical image analysis or synthesis task, and it can be involved with real, quaternion, or hypercomplex-valued models, generalizing their adoption to single-channel data. QUAVE first extracts different sub-bands through the quaternion wavelet transform, resulting in both low-frequency/approximation bands and high-frequency/fine-grained features. Then, it weighs the most representative set of sub-bands to be involved as input to any other neural model for image processing, replacing standard data samples. We conduct an extensive experimental evaluation comprising different datasets, diverse image analysis, and synthesis tasks including reconstruction, segmentation, and modality translation. We also evaluate QUAVE in combination with both real and quaternion-valued models. Results demonstrate the effectiveness and the generalizability of the proposed framework that improves network performance while being flexible to be adopted in manifold scenarios and robust to domain shifts. The full code is available at: \url{https://github.com/ispamm/QWT}.
\end{abstract}


\section{Introduction}
\label{sec:intro}



The generalizability challenge in neural network deployment is still a wide-open problem, being amplified by the increasing number of available image datasets coming from different sources and diverse acquisition processes. This matter becomes crucial when dealing with medical image datasets, whose samples are generated by a variety of hospitals, devices, and operators, thus similar datasets can indeed have wide variations. Data of the same medical exam obtained in diverse hospitals may suffer from a domain shifting affecting the generalizability of pretrained models that often have to be re-trained from scratch on the new data \cite{Zhou2022ECCV}. In fact, deep learning techniques have profoundly permeated the field of medical image analysis \cite{ZHOU2023106959, dhar2023, KAZEROUNI2023102846} 
reaching impressive results in manifold applications, including detection and classification \cite{Lyu2022Alzheimer, 10218990, DAI2023109108}, segmentation \cite{Tomar2022TNNLS, Ma2024}, reconstruction, and inverse problems solution \cite{song2022solving, Selivanov2023, Chen2021TarGANTG}, among others. However, such approaches are usually task- or data-specific, thus losing generalizability and undermining their adoption in different contexts. Furthermore, medical images are usually high-quality and full-of-details images that have to be resized and downscaled to be handled by neural networks due to computational limits. This process may compromise image quality and cause a loss of crucial details even in the region of interest, which is often a tiny portion of the whole picture. Due to such image daintiness, the preprocessing often plays a pivotal role in medical analysis \cite{Clough2020PAMI, SAIFULLAH20233021}, being a mandatory step before developing any deep learning model for image datasets \cite{Song2021DoesPH}. 

As part of preprocessing, techniques for identifying important and explanatory training examples have been developed for natural images \cite{Paul2021NeurIPS, Barshan2020RelatIFIE} aiming at discarding superfluous data samples and improving the generalization ability of neural networks. Similarly, several works try to establish the most representative patches (or whole sample) to be involved in the training of medical deep learning approaches showing the ability to improve models performance \cite{Shen2022TMI, Hanik2022BIB}. Simultaneously, a further approach consists of fusing different image modalities in a single and more informative sample \cite{ZHANG2023104545, Tamg2021ICMIPE}. This method is particularly suitable when the dataset contains multimodalities or multiresolution samples. Fusing different modalities combines important information coming from the various acquisitions and may build a more detailed structure, improving pathology examination accuracy. A common approach for medical image fusion aims at leveraging wavelet transforms that split the original image into different frequency components, providing different levels of detail \cite{Yang2010, Yan2012wavelet, 2023waveletmedical}. Recently, a further improvement has been brought by the adoption of the quaternion wavelet transform (QWT) that provides a hypercomplex approach with additional sub-bands and several advantages, including shift invariance, over the standard real-valued wavelet \cite{Zhancheng2020, Ngo20221, Chai2017ImageFU, grouse10268972}. 

\begin{figure*}
    \centering
    \includegraphics[width=0.85\textwidth]{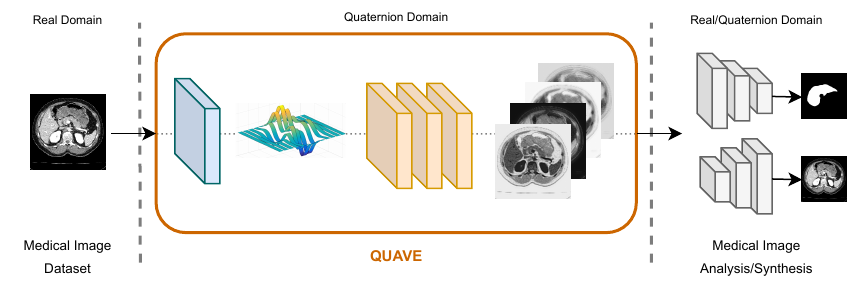}

    \caption{Illustration of the QUAVE framework. Any medical image dataset can be processed in the quaternion domain via QUAVE, which carries out the QWT and the neural feature extraction. The training procedure can be then performed either in the real or in the quaternion domain depending on the user choice, producing a real-valued output. We integrate a quaternion-based framework in the analysis and synthesis of real-valued medical image datasets.}
    \label{fig:framework}
\end{figure*}

In this paper, we introduce a novel generalizable framework for medical image analysis and synthesis, referred to as quaternion wavelet network (QUAVE). The proposed approach is data-agnostic, and it does not suffer from domain shifting in data, thus it can be employed with any kind of medical image dataset, for any small and large-scale neural model, and for different tasks. 
We integrate a quaternion-based preprocessing in a real-valued analysis procedure, to leverage quaternion capabilities in any real-valued scenario, while simultaneously further generalizing the adoption of quaternion and hypercomplex neural networks.
QUAVE is built upon two core blocks. The first block exploits quaternion wavelet potentials to extract salient features from medical images that are usually full of details and require detailed preprocessing to preserve crucial content characteristics. These features are then processed by the second QUAVE block, where a neural model extracts the most informative content from the QWT bands to build enhanced representations that can be involved as input to any model for image analysis. Indeed, thanks to the QWT low-frequency and high-frequency sub-bands, we make any model for analysis aware of both approximations and details of the input image, improving its ability to reconstruct or generate images full of details, while also enhancing segmentation capabilities. The overall structure of our framework is depicted in Fig.~\ref{fig:framework}. QUAVE output can be generalized to other datasets without requiring the repetition of the training procedure.
Furthermore, since the proposed wavelet-based preprocessing is performed in the quaternion domain, we can easily seize quaternion-valued neural networks (QNNs) for the learning stage to further exploit the capabilities of hypercomplex algebra. Indeed, QNNs have been proven to achieve interesting results in processing images \cite{SPMDemistHyper, 10122661, Zhang2022TNNLS, Jia2022TIP, Lin_2023_ICCV, grassucci2022, LOPEZ2024140}.
Due to the four-dimensional nature of quaternions, QNNs properly handle inputs with $4$ dimensions/channels, thus restricting the use of QNNs to a few sets of data. 
Instead, thanks to QUAVE, we generalize the usage of quaternion and hypercomplex neural networks for single-channel inputs while also improving performance. In fact, by exploiting the proposed approach it is possible to easily handle grayscale and one-channel images in QNNs. Indeed, through the QUAVE processing, we obtain four sub-bands that can be easily encapsulated in a quaternion. Due to the obvious correlations among these sub-bands, QNNs equipped with our method can improve performance by grasping such relations and sharing information among channels.

We prove our theoretical claims through a wide experimental evaluation covering three different tasks and three common medical image datasets. 
In the largest part of the test we conduct, QUAVE enhances model performance outperforming previous techniques, according to a plethora of objective metrics and a visual inspection. Therefore, the proposed approach is flexible enough to be adopted with any kind of dataset, especially if it contains images full of details that can be exploited by high-frequency sub-bands.

The rest of the paper is organized as follows. Section~\ref{sec:qalg} introduces basic concepts of quaternion algebra, Section~\ref{sec:qrepr} expounds on the wavelet and quaternion wavelet transforms for images, and Section~\ref{sec:select} presents the proposed selection method. The experimental setup is then introduced in Section~\ref{sec:exp_setup} and validated in Section~\ref{sec:exp}. Finally, conclusions are drawn in Section~\ref{sec:con}.

\section{Fundamentals of Quaternion Learning}
\label{sec:qalg}
In this section, we expound on basic concepts of quaternion algebra, the foundations of quaternion neural networks and how they can be involved in a variety of studies. Indeed, real-valued data can be easily processed with quaternion-valued models by encapsulating input dimensions into the components of a quaternion. Similarly, a quaternion framework can be handily integrated into a real-valued analysis.

\subsection{Quaternion Algebra}

The quaternion set of numbers $\bH$ belongs to the class of Clifford algebras, being a four-dimensional associative, normed, and non-commutative division algebra over real numbers. While complex numbers are defined by a real-valued component and one imaginary unit, quaternion numbers comprise two additional imaginary components, being defined as:

\begin{equation}
\label{eq:q}
    q = q_0 + q_1 \ii + q_2 \ij + q_3 \ik,
\end{equation}

\noindent whereby $q_c, c \in \{0,1,2,3\}$ are real-valued coefficients and $\ii, \ij, \ik$ imaginary units, which comply with the properties: 
\begin{enumerate}
    \item $\ii^2 = \ij^2 = \ik^2 = -1$;
    \item $\ii \times \ij = \ik, \ij \times \ik = \ii, \ik \times \ii = \ij$;
    \item $\ii \times \ij \neq \ij \times \ii, \ij \times \ik \neq \ik \times \ij, \ik \times \ii \neq \ii \times \ik$,
\end{enumerate}

\noindent where $\times$ is the vector product in $\bR^3$. An alternative quaternion representation can be derived considering the magnitude and the three phases:

\begin{equation}
\label{eq:qmagphase}
    q = |q| e^{\ii \theta_1} e^{\ij \theta_2} e^{\ik \theta_3},
\end{equation}

\noindent in which $|q|$ is the amplitude of the quaternion and $(\theta_1, \theta_2, \theta_3) \in [ \pi, \pi) \times [-\pi/2, \pi/2) \times [-\pi/4, \pi/4]$ are the phase angles of it. The latter can be computed as follows:

\begin{equation}
\label{eq:phases}
    \begin{cases}
    \theta_1 = \arctan \left( \frac{2q_0 \cdot q_2 + q_1 \cdot q_3}{q_0^2 + q_1^2 - q_2^2 - q_3^2} \right) \\
        \theta_2 = \arctan \left( \frac{q_0 \cdot q_1 + q_2 \cdot q_3}{q_0^2 - q_1^2 + q_2^2 - q_3^2} \right) \\
        \theta_3 = \frac{1}{2} \arctan \left( 2 \left( q_0 \cdot q_3 - q_3 \cdot q_1 \right) \right).
    \end{cases}
\end{equation}

The quaternion domain is endowed with a norm $|q| = \sqrt{q_0^2+q_1^2+q_2^2+q_3^2+q_4^2}$, a conjugate $q^* = q_0 - q_1 \ii - q_2 \ij - q_3 \ik$, the operations of scalar multiplication $q \cdot p = q_0p_0 + q_1p_1\ii + q_2p_2\ij + q_3p_3\ik$, and associative multiplication of elements.

Due to the vector product non-commutative property, the Hamilton product has been introduced for properly modelling imaginary unit interplays \cite{Ward1997}:

\begin{equation}
\label{eq:qprod}
    q \otimes p = \left[ {\begin{array}{*{20}c}
   \hfill {{q}_0 } & \hfill { - {q}_1 } & \hfill { - {q}_2 } & \hfill { - {q}_3 } \\
   \hfill {{q}_1 } & \hfill {{q}_0 } & \hfill { - {q}_3 } & \hfill {{q}_2 } \\
   \hfill {{q}_2 } & \hfill {{q}_3 } & \hfill {{q}_0 } & \hfill { - {q}_1 } \\
   \hfill {{q}_3 } & \hfill { - {q}_2 } & \hfill {{q}_1 } & \hfill {{q}_0 } \\
\end{array}} \right] \left[ {\begin{array}{*{20}c}
   {p_0 } \hfill  \\
   {p_1 } \hfill  \\
   {p_2 } \hfill  \\
   {p_3 } \hfill  \\
\end{array}} \right].
\end{equation}

\subsection{Learning in the Quaternion Domain}

Quaternion neural networks (QNNs) are built upon the Hamilton product for multiplying and convolving weights with input data \cite{ParcolletAIR2019, ArenaISCAS1994}.
More precisely, a real-valued fully connected layer with weights matrix $\mathbf{W}$ and bias $\mathbf{b}$ has the following expression:

\begin{equation}
    \mathbf{y} = \mathbf{W}\mathbf{x} + \mathbf{b},
\end{equation}

\noindent where the real-valued input $\mathbf{x}$ is modelled into the real-valued output $\mathbf{y}$.
A quaternion fully connected (QFC) layer involves quaternion characters and its definition is based on the Hamilton product between the weights matrix, arranged in the form of \eqref{eq:qprod}, and the quaternion input as:

\begin{equation}
\label{eq:qfc}
    \mathbf{y} = \mathbf{W} \otimes \mathbf{x} + \mathbf{b}.
\end{equation}

The QNNs formulation through the Hamilton product has two key advantages. First, since the quaternion weight matrix is composed of four sub-matrices $\mathbf{W}_c$, with  $c\in\{0,1,2,3\}$, each with $1/16$ the parameters of the complete matrix $\mathbf{W}$, and these sub-matrices are reused to build the final weight matrix $\mathbf{W}$ according to \eqref{eq:qprod}, QNNs save $75\%$ of free parameters with respect to real-valued counterparts. Second, due to such sharing of weight sub-matrices, each parameter is multiplied by each dimension of the input (e.g., by each channel of RGB images, or of multichannel signals), thus capturing complex relationships among input dimensions and preserving their correlations \cite{ParcolletICASSP2019a, Grassucci2021Entropy}. This allows QNNs to gain comparable results when processing multidimensional data despite the lower number of free parameters. However, due to the four-dimensional nature of quaternion numbers, QNNs are mainly limited to multidimensional signals, while their extension to one-channel and monodimensional data is still an open problem.

\section{Quaternion Wavelet Representation of Medical Images}
\label{sec:qrepr}

\begin{figure}[t]
    \centering
    \includegraphics[width=\linewidth]{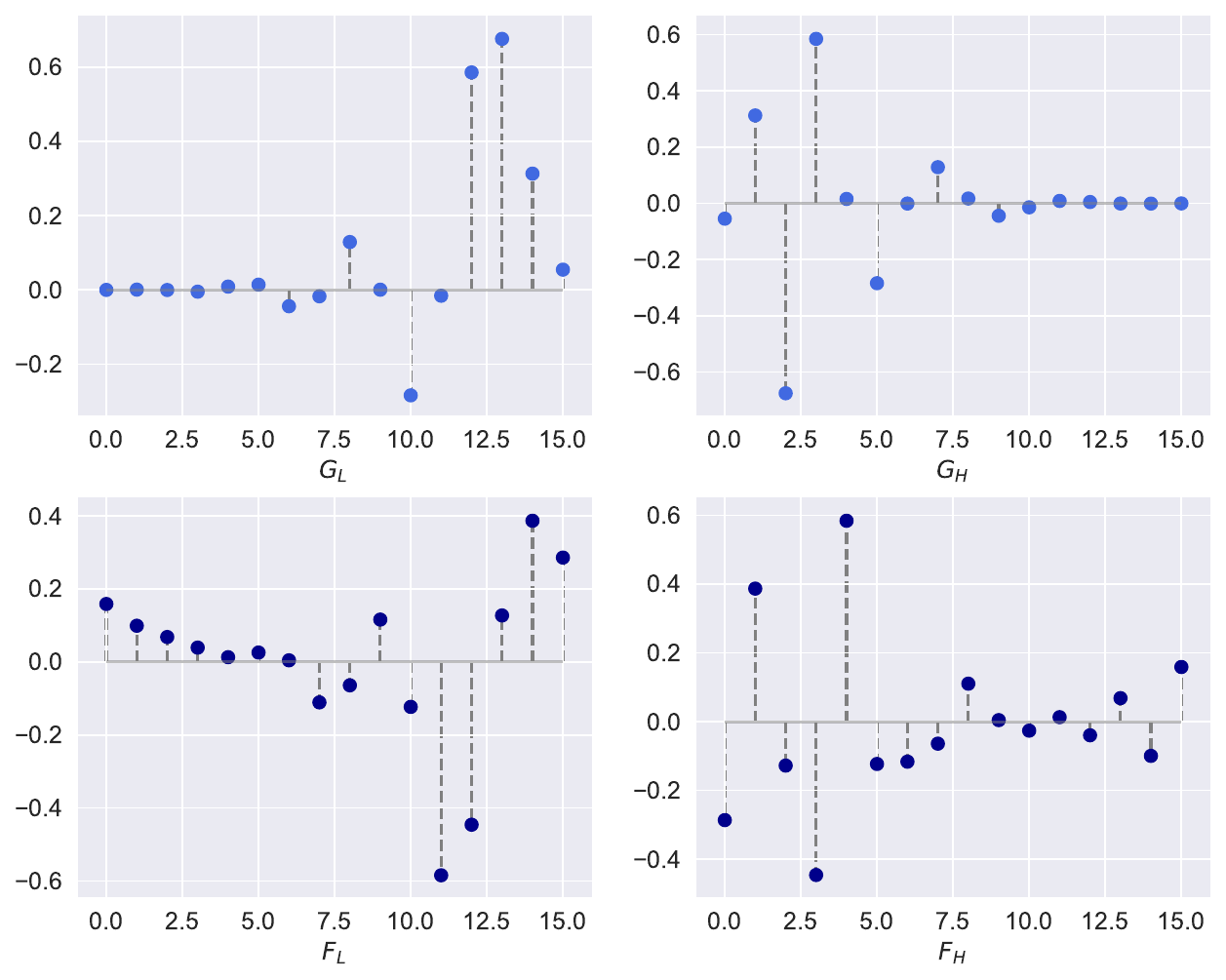}
    \caption{Filters for QWT: $G_L$ and $G_H$ are extracted from the wavelet db8, while $F_L$ and $F_H$ are Hilbert transforms of the previous ones.}
    \label{fig:filters}
\end{figure}

In this section, we expound on the proposed approaches for extracting salient characteristics from input images. Moreover, we also explain how to leverage these features with quaternion neural networks.

\subsection{Quaternion Embedding of Real-Valued Wavelet Representations}
The 1D discrete wavelet transform (1D-DWT) characterizes a signal $f(t)$ by shifting a scaling function $\phi(t)$ and shifting and scaling a wavelet function $\psi(t)$ \cite{IntroWave1992}. In case of 2D signals, the 2D-DWT can be obtained by involving tensor products of 1D-DWTs over the two dimensions, resulting in the scaling function $\phi(x)\phi(y)$ and three wavelets $\psi(x)\psi(y)$, $\phi(x)\psi(y)$, and $\psi(x)\phi(y)$ that extract diagonal, horizontal and vertical features, respectively \cite{Vetterli1992, Chan2004QWT}. Through this decomposition, the 2D-DWT (from now on, DWT for simplicity) of an image allows for the derivation of a low-frequency (LL) sub-band and three high-frequency (LH, HL, HH) sub-bands. However, this kind of decomposition has some disadvantages. First of all, it lacks translation invariance. Secondly, it does not contain image phase information that usually cares about describing the spatial information of image contents. Therefore, even small shifts in the image content, which are quite common in datasets acquired through different devices or by diverse operators, can crucially affect wavelet magnitude \cite{Selesnick2005TSP, Chan2008Coherent}. Indeed, phase information may be crucial in medical images that could contain artifacts due to patient movements or inaccurate equipment. Consequently, in order to build a more powerful representation of DWT image features, we propose to leverage quaternion algebra properties by encapsulating the four sub-bands in a quaternion as $q=\text{LL}+\text{LH} \ii +\text{HL} \ij +\text{HH} \ik$ and then exploiting its magnitude and phases characterization in \eqref{eq:qmagphase}. The phase angles can be then computed following \eqref{eq:phases}. In this way, the input image is first decomposed into different frequency sub-bands by means of a DWT and then processed through quaternion algebra to get also the missing phase information, building a more accurate representation. Such manipulated data can be then fed into real, quaternion, or hypercomplex-valued neural networks without needing additional preprocessing.


\subsection{Enhanced Representation via Quaternion Wavelet Transform}
\label{subsec:qwt}
\begin{figure*}
    \centering
    \includegraphics[width=\linewidth]{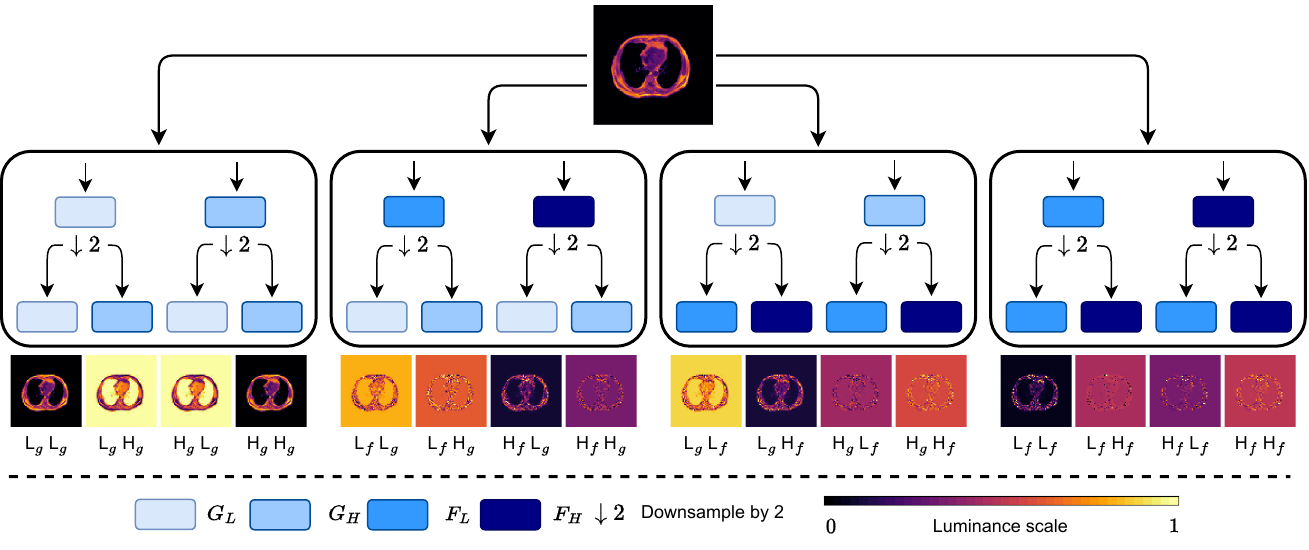}
    \caption{Dual-tree quaternion wavelet transform (QWT). The input image is decomposed in sixteent sub-bands by interleaving low-pass and high-pass filters.}
    \label{fig:qwt}
\end{figure*}
An enhanced representation of medical images, that we embed in our QUAVE framework, involves the use of the dual-tree quaternion wavelet transform (QWT) \cite{Chan2004QWT}. It comprises four quaternion wavelet transforms that are built by means of a real DWT and its three Hilbert transforms on the $x,y$ and $xy$ axis. The QWT decomposes the input image in four quaternion wavelet sub-bands for a total of 16 real sub-bands. As for the DWT, the transformation result comprises a scaling function $\phi_q$ and three wavelets $\psi^D_q, \psi^V_q, \psi^H_q$ in the vertical, horizontal and diagonal direction, respectively \cite{Zhang2020QWT, Chan2008Coherent}. More formally, they are defined by:

\scriptsize
\begin{equation}
\label{eq:qwt_phases}
\begin{aligned}
\phi_q&=\phi_g(x) \phi_g(y)+\phi_f(x) \phi_g(y) \hat{\imath}+\phi_g(x) \phi_f(y) \hat{\jmath}+\phi_f(x) \phi_f(y) \hat{\kappa} \\
\psi_q^V&=\psi_g(x) \phi_g(y)+\psi_f(x) \phi_g(y) \hat{\imath}+\psi_g(x) \phi_f(y) \hat{\jmath}+\psi_f(x) \phi_f(y) \hat{\kappa} \\
\psi_q^H&=\phi_g(x) \psi_g(y)+\phi_f(x) \psi_g(y) \hat{\imath}+\phi_g(x) \psi_f(y) \hat{\jmath}+\phi_f(x) \psi_f(y) \hat{\kappa} \\
\psi_q^D&=\psi_g(x) \psi_g(y)+\psi_f(x) \psi_g(y) \hat{\imath}+\psi_g(x) \psi_f(y) \hat{\jmath}+\psi_f(x) \psi_f(y) \hat{\kappa}
\end{aligned}
\end{equation}
\normalsize

\noindent whereby $g, f$ are a real-valued filter and its corresponding Hilbert transform.

More concretely, the QWT of an image can be computed by combining four filters. To this end, an asymmetric, orthogonal and biorthogonal Daubechies wavelet with 8 vanishing moments (db8) can be considered, following \cite{Ngo2022}. From the db8, the decomposition low-pass ($G_L$) and high-pass filters ($G_H)$ are extracted, and then the Hilbert transform is applied to them to obtain the filters $F_L$ and $F_H$. A plot of such filters is shown in Fig.~\ref{fig:filters}. Once $G_L, G_H, F_L$ and $F_H$ are available, the QWT is calculated by combining filters and downsampling operations according to Fig.~\ref{fig:qwt}.
The result comprises $4$ low-frequency and approximation sub-bands ($\phi_q$) and $12$ high-frequency sub-bands ($\psi^V_q, \psi^H_q, \psi^D_q$). Furthermore, the four components in \eqref{eq:qwt_phases} can be organized into a quaternion, whose square magnitude is non-oscillatory, meaning that the transform is approximately shift-invariant \cite{Chan2004QWT}.

Among these featured images, some previous works suggest only involving the four low-frequency components as input to the model since, under an empirical evaluation, they improve classification accuracy \cite{Ngo2018lowcoeff, Zou2007IJCNN}. However, these tests were conducted with vanilla classifiers without involving deep learning approaches that are notoriously more powerful. Indeed, shallow models may not have enough capacity to handle very complex inputs such as high-frequency wavelet details. Neural networks, instead, are more complex systems and more detailed inputs can further enhance model expressiveness, improving performance. Furthermore, for the task of image classification, there may be sufficient to have an approximate representation of the input, while a lack of high-frequency details may undermine results on other tasks such as reconstruction, segmentation or generation. On the other hand, high-frequency features are less informative with respect to low ones, thus a proper combination of low and high sub-bands can better represent input images. However, involving the sixteen sub-bands as input to a neural model may undermine its performance due to redundant features, forcing the model to distinguish between relevant and uninformative channels before effectively learning a proper representation.
Furthermore, QNNs require four-channel inputs to be encapsulated into a quaternion, thus it is not suitable to directly involve the sixteen sub-bands.

To this end, the QUAVE framework involves a selection strategy, detailed in the next section, to choose the most explanatory features to be fed as input to real and quaternion-valued networks.
\begin{figure*}[t]
    \centering
    \includegraphics[width=0.9\linewidth]{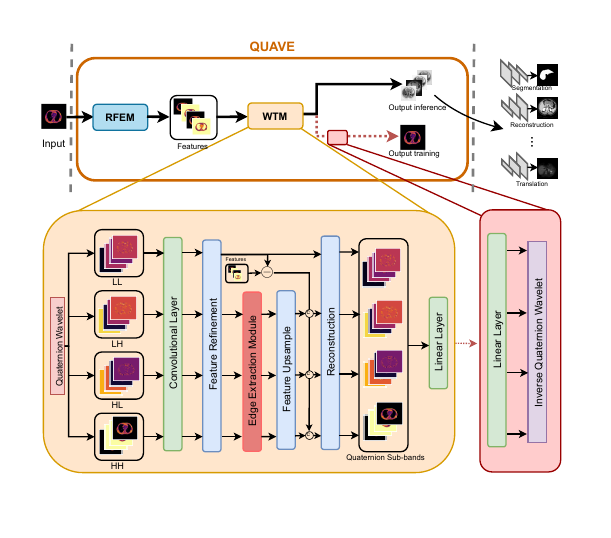}
    \caption{Detailed QUAVE architecture.
    Being a quaternion-based generalizable framework, the WTM enhances neural model performance when processing rich-of-detail real images, such as medical ones. Along with the inverse quaternion wavelet block, the WTM performs a salient feature extractor by leveraging quaternion algebra properties to achieve more detailed insights. During inference, the output of our framework can be then involved in any neural model either in the real or quaternion domain, improving performance in a variety of tasks (e.g., segmentation, reconstruction, and generation, among others).}
    \label{fig:frameworkfusione}
\end{figure*}
\section{The Proposed QUAVE}
\label{sec:select}

In this section, we expound upon our pathway for extracting the most informative representation from the sixteen features obtained through the Quaternion Wavelet Transform (QWT). While the Discrete Wavelet Transform (DWT) produces four sub-bands that can be directly utilized as input for models, the QWT generates sixteen sub-bands, potentially resulting in redundant information if they are directly used as input to the network. Moreover, quaternion-valued models require exactly four-channel inputs, rendering the direct conveyance of the sixteen QWT sub-bands to the network impractical.
To address these challenges, we propose a learning-based approach to build the most informative representation among the sixteen quaternion wavelet sub-bands for involving it as enhanced input to neural models.
We build QUAVE upon the recently demonstrated effectiveness of wavelets in various tasks, such as image fusion methods \cite{Chai2017ImageFU} involving multimodal image fusion using wavelets and multiple features. Among these methods, wavelets are particularly effective in enhancing super-resolution by enriching output images with improved fine-grain details \cite{dastmalchi2022super, aloisi2024waveletdiffusionganimage, zou2022joint}.
For this reason, we propose a neural model that leverages quaternion wavelets to perform super-resolution, extracting the most informative content by means of the quaternion wavelet features. We propose to leverage transfer learning, wherein QUAVE is trained to perform a super-resolution task on a generic medical dataset (e.g., IXI) and then involved in downstream tasks with the knowledge it has acquired during training.

\subsection{Pretraining}
We pretrain the model with the objective of enhancing resolution by reconstructing high-quality images with enriched detail through the quaternion wavelet and feature refinement techniques like those in \cite{zou2022joint}.
QUAVE pretraining combines a content loss $\mathcal{C}_{loss}$ and a wavelet-based loss $\mathcal{W}_{loss}$ component to enforce finer structure and detail preservation in the predictions of the model. The content loss is an L1 loss between the generated image and the ground truth image, which penalizes direct differences in pixel values, while the wavelet loss captures differences in frequency information.
This is because the ground truth image $x_{gt}$ and the prediction $x_{pred}$ are transformed into the wavelet domain. The wavelet loss is computed as the L1 loss between these transformed representations $wavelet_{gt}$ and $wavelet_{pred}$.
The final loss function combines the content and wavelet losses, with $\alpha$ controlling the influence of the wavelet loss.

\begin{equation}
\mathcal{L} = \mathcal{C}_{loss} + \alpha \cdot \mathcal{W}_{loss}
\end{equation}

Once reached the convergence, the pretrained QUAVE model is utilized to extract the features for a downstream task of interest, such as reconstruction, segmentation, or image translation. Extracting the features entails employing the pretrained QUAVE model and, as depicted in Fig.~\ref{fig:frameworkfusione}, selecting the weights from the second-last layer, prior to the inverse wavelet transform.

\begin{figure}
    \centering
    \includegraphics[width=\linewidth]{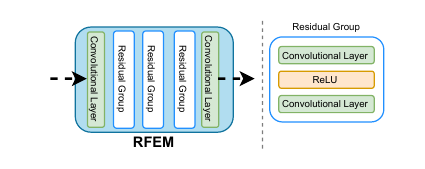}
    \caption{The Rough Feature Extraction Module (RFEM) is responsible for initial feature extraction before the Quaternion Wavelet Transform (QWT). It consists of a convolution layer and multiple residual groups designed to extract both shallow and deep features from the input image and does not apply QWT. This prepares the data with enriched initial features before entering the QWT, making the wavelet transform more effective.}
    \label{fig:rfem}
\end{figure}

\subsection{Architecture}
The architecture of QUAVE consists of two principal modules, as shown in Fig.~\ref{fig:frameworkfusione}. The first one is the rough feature extraction module (RFEM), displayed in Fig.~\ref{fig:rfem}, which provides the necessary features for performing the quaternion wavelet transform. The residual groups in RFEM are structured like in \cite{zhang2018image} and visualized in Fig.~\ref{fig:rfem}. Note that to improve the clarity of the figure, the residual connections among the layers are not reported. Subsequently, the wavelet transform module (WTM) takes the features extracted by the RFEM and applies the QWT to those. Moving forward, the module operates in four groups: one low-frequency group and three high-frequency groups. Before entering the feature refinement block, we reduce the dimensionality of the image features by a factor of four using a convolutional layer. 
Inside the WTM, the Edge Extraction Module focuses on high-frequency information and captures essential edge details of the image. The module utilizes all the high-frequency sub-bands (HL, LH, HH) to create an edge feature map (EFM), which consolidates all high-frequency information to enhance edge detail. The EFM is computed with the L2 Norm:
\begin{equation}
\text{EFM} = ||F_x||_2 = \sqrt{|F_{LH}|^2 + |F_{HL}|^2 + |F_{HH}|^2}
\end{equation}
where \( F_x \) represents the high-frequency sub-bands (\( F_x \in \{F_{LH}, F_{HL}, F_{HH}\} \)), and \(\text{EFM}\) denotes the edge feature map.
Next, the edge information is incorporated back into each frequency sub-band through concatenation.

Two linear layers are inserted before the inverse quaternion wavelet transform to extract the enriched features of the sub-bands, which outputs the final super-resolution image.
Those layers allow us to retrieve the desired features from QUAVE, representing four sub-bands of the quaternion wavelet, that are considered the most informative by the neural model, as they activate the neurons that yield the best performance.

During the training phase of a downstream task of interest (e.g. segmentation), the original data flows into the pretrained QUAVE that performs inference and outputs the four most representative sub-bands. This enhanced representation is then fed as input to the downstream model to perform the task, as shown in the upper part of Fig.~\ref{fig:frameworkfusione}. 
Interestingly, the features representing the four sub-bands of the quaternion wavelet can naturally represent a quaternion number, and this perfectly fits on quaternion-valued neural networks.

\section{Experimental Setup}
\label{sec:exp_setup}

We perform experiments in multiple tasks and with different benchmark datasets to show the generalizability of our framework and to strengthen our theoretical claims. Specifically, we utilize QUAVE pretrained on the IXI dataset for downstream tasks involving grayscale medical image datasets and QUAVE pretrained on the Kvasir dataset for downstream tasks on RGB medical image datasets.
We first consider tasks regarding image analysis such as reconstruction and segmentation and then an image synthesis task, that is, image modality translation with a large-scale model. We conduct analysis on two algebraic domains: first, we set up experiments to prove that our framework can enhance the performance of real-valued neural networks.
When using the term \textit{real-valued}, we are referring to neural networks defined in the domain of real numbers, therefore not relying on quaternion numbers and thus not involving quaternion-valued layers, to which instead we refer as \textit{quaternion-valued}.
Second and most importantly, we perform evaluations with quaternion-valued models to show that QUAVE outperforms other preprocessing techniques and it is flexible to be adopted in different domains, further improving QNNs generalizability.
Moreover, we consider diverse types of input images. We take into account medical grayscale and RGB images, and multimodal images, proving that the proposed method can be generalized to any kind of input, thus being data-agnostic and not suffering from domain shifting.

\subsection{Description of the Tasks}
\subsubsection{Reconstruction}
As a first step of our evaluation, we take into account the IXI dataset\footnote{\url{https://brain-development.org/ixi-dataset}} for the reconstruction task. It comprises magnetic resonance images (MRI) collected from three different hospitals in London, and it is organized as follows: T1, T2, PD-weighted images, MRA images and diffusion-weighted images. We consider the combination of the multimodal T1 and T2 images of $581$ and $578$ patients respectively.
Each 3D MRI scan is stored in NIFTI format and is composed of a stack of 2D slices that together represent the full volumetric structure of the brain. For example, a typical T1-weighted volume comprises approximately 130–150 axial slices, while a T2-weighted volume consists of around 112–136 axial slices. In our experiments, for the purpose of the reconstruction task, we extract a single, representative 2D slice per patient. This approach simplifies the reconstruction task by not processing the entire 3D volume or a large number of slices.

After this step, we obtain $1156$ images from the combination of T1 and T2 ones, each of size $256\times256$. We employ $866$ of them for training and $290$ for validation.  

\subsubsection{Segmentation}
The second set of experiments comprises the segmentation task on the Kvasir-SEG dataset \cite{kvasirseg}. This dataset contains snapshots of polyps with manually annotated masks verified by an experienced gastroenterologist. The $1000$ samples range from $332\times487$ to $1920\times1072$ pixels, and we resize them to $256\times256$. We involve $750$ samples for training and the remaining part for validation. Interestingly, and differently from previous datasets, the Kvasir-SEG dataset contains RGB images with three channels. Therefore, we train the real-valued model with original three-channel samples, while we apply the DWT and the QWT computed on the corresponding grayscaled image.
\subsubsection{Image Modality Translation}
The third set of experiments concerns image modality translation on the CHAOS - Combined (CT-MR) Healthy Abdominal Organ Segmentation dataset. It contains CT and MR scans from unpaired abdominal image series, the ground truth is generated with the auxiliary participation of multiple experts' annotations and introduced for the first time in \cite{CHAOSdata2019}. The data comes from $80$ patients from the Department of Radiology of Dokuz in Turkey, half of them have done only a single CT scan while the others went through MR scans.
For our experiments we have used all the data resizing the slices at $128 \times 128$, obtaining approximately $4144$ samples, we randomly select $80\%$ from all the modalities to be the training dataset, and the remaining as test one. 

\begin{figure*}[t]
    \centering
    \includegraphics[width=0.8\linewidth]{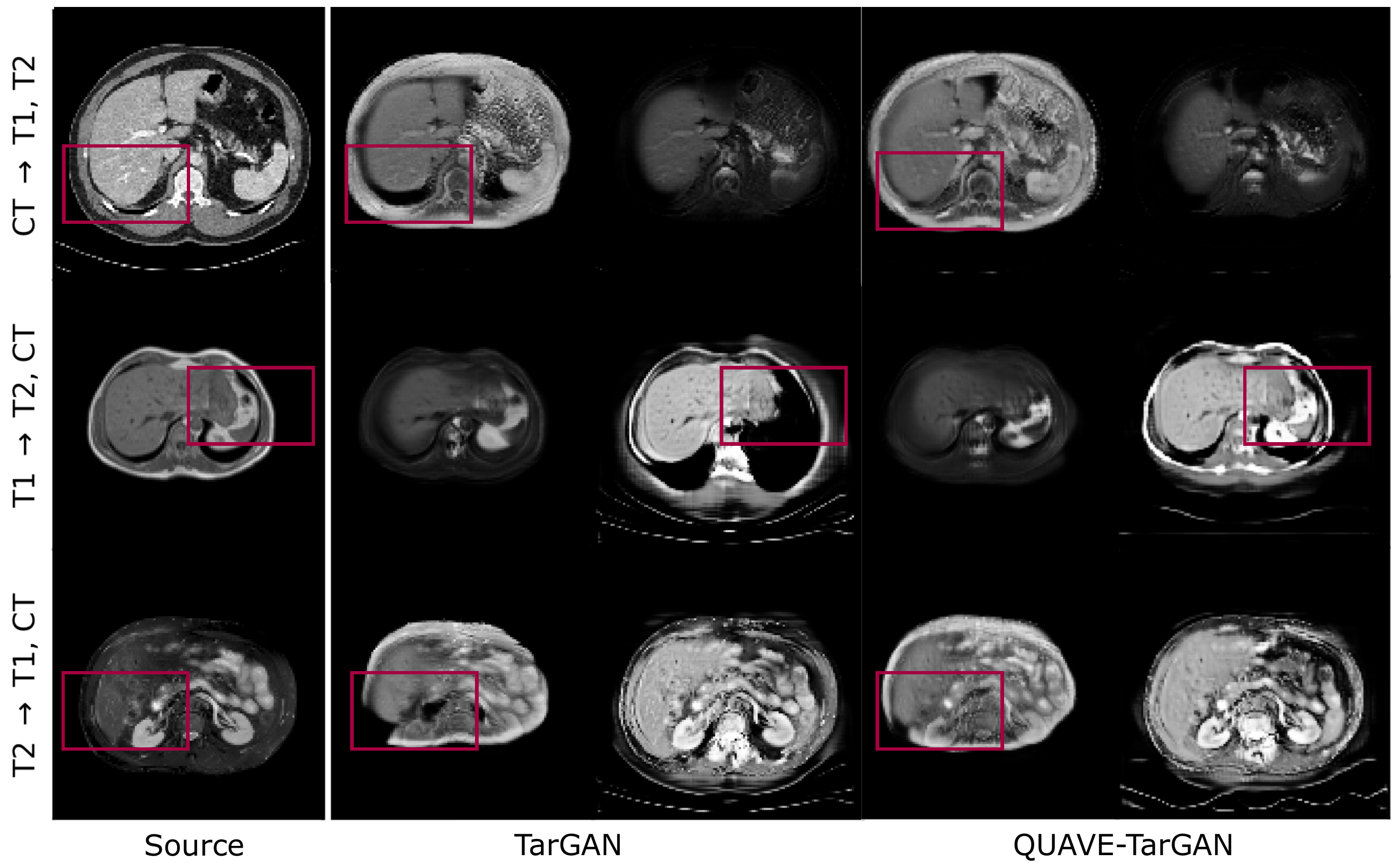}
    \caption{Random samples of TarGAN image modality translations on the CHAOS dataset. The first column refers to the original domain images, while the others refer to the corresponding translation in the other domains. 
    }
    \label{fig:targan_results}
\end{figure*}

\subsection{QUAVE in Combination with Real- and Quaternion-Valued Models}

We involve different models in our tests. For the reconstruction task, we consider a vanilla convolutional autoencoder (CAE) \cite{ParcolletICASSP2019a}, whereby the encoder is composed of three convolutional layers $[28, 64, 128]$ with kernels $[8, 3, 4]$ for the IXI dataset. The decoder has a mirrored structure with transposed convolutions. 
The quaternion CAE (QCAE) \cite{ParcolletICASSP2019a} has the same backbone while involving quaternion convolutions instead of real-valued ones. For each reconstruction experiment, we set the learning rate of the Adam optimizer to $0.0001$, the batch size to $4$, and the number of epochs to $10$, repeating the learning process for multiple runs with different seeds.

Concerning the segmentation task, we consider a U-Net \cite{Ronneberger2015UNET} and a more advanced U-Net++ \cite{Zhou2020UNetRS} in our experiments, as they are widespread models for this scenario.
In addition, quaternion counterparts for these architectures are available and have been validated in prior work, which enables us to perform a direct and fair comparison between real-valued and quaternion-valued approaches within our QUAVE framework. Our primary focus is on demonstrating the generalizability and benefits of integrating QUAVE with existing segmentation networks. Thus, using U-Net and U-Net++ allows us to clearly highlight the improvements brought by our framework rather than solely pursuing the highest possible segmentation accuracy.
The U-Net encoder comprises four blocks, each with two convolutions, batch normalization, and ReLU activation function, while max pooling is interleaved between each block. The number of filters increases going deeper, being $[32, 64, 128, 256]$, each with kernel sizes $3$ and stride $1$. The bottleneck adds a further block with $512$ filters. Then, the information is passed to the decoder, which has the same architecture as the encoder except for the replacement of transposed convolutions instead of max pooling operations. A final convolutional layer refines the resulting segmentation map with $1$ channel. Similar to the QCAE, the quaternion U-Net (QU-Net) replaces convolutions and transposed convolutions with quaternion-valued counterparts. The more advanced U-Net++ has a similar structure to the original one, inserting however dense connections that aggregate intermediate feature maps. Every hyperparameter is set according to the paper \cite{Zhou2020UNetRS}. In these experiments, as for the reconstruction task, we set the learning rate to $0.0005$ for the U-Net and to $0.0003$ for the U-Net++, as suggested in the paper \cite{Zhou2020UNetRS}, with Adam optimizer and the batch size to $4$. We train all the models for $50$ epochs, repeating the learning process for multiple runs with different seeds.
\begin{table*}[t]
\centering
\caption{Results for task 3 image modality translation on the CHAOS dataset (Segmentation metrics) for real and quaternion-valued models.}
 \resizebox{\textwidth}{!}{%

\begin{tabular}{lcccc} 
\toprule
Model  & DSC$\uparrow$ & S-Score$\uparrow$ & mIoU$\uparrow$ & MAE$\downarrow$ \\
\midrule
TarGAN \cite{Chen2021TarGANTG}  & $87.53 \pm 0.6474$ & $76.23 \pm 9.5988$ & $0.7812 \pm 0.0090$ & $0.0489 \pm 0.0007$ \\
QUAVE-TarGAN  & $\mathbf{90.53 \pm 4.3608}$ {(+3.4\%)} & $\mathbf{84.41 \pm 4.2848}$ {(+10.7\%)} & $\mathbf{0.8320 \pm 0.0676}$ {(+6.5\%)} & $\mathbf{0.0487 \pm 0.0004}$ {(+0.4\%)} \\
\midrule
QTarGAN & $\mathbf{83.56 \pm 1.1758}$ & $76.85 \pm 1.2373$ & $0.7246 \pm 0.0158$ & $0.0492 \pm 0.0011$ \\
QUAVE-QTarGAN  & $83.02 \pm 0.7664$ {($-0.6\%$)} & $\mathbf{78.90 \pm 1.1821}$ {(+2.6\%)} & $\mathbf{0.7360 \pm 0.0096}$ {(+1.6\%)} & $\mathbf{0.0488 \pm 0.0005}$ {(+0.8\%)} \\
\bottomrule
\end{tabular}
}
\label{tab:targan_results_dsc_s_miou_mae}
\end{table*}

Regarding the image modality translation task, we consider the recent Target-Aware Generative Adversarial Network (TarGAN) \cite{Chen2021TarGANTG}. The TarGAN translates images from a source domain to a target domain while paying particular attention to the region of interest passed as a segmentation map during the translation.
In the TarGAN framework, segmentation information is explicitly incorporated into the I2I translation process. Specifically, given an image \(x_s\) from the source modality \(s\) and its corresponding target area label \(y\), a binarization operation \(y \cdot x_s\) is applied to isolate the target area, yielding \(r_s\), an image containing only the target region. TarGAN then translates both the full input image \(x_s\) and the extracted target area \(r_s\) into their corresponding representations in the target modality \(t\), denoted as \(G(x_s, r_s, t) \rightarrow (x_t, r_t)\).
To accomplish this, TarGAN employs a double-stream generator with a shared middle block and two pairs of encoder-decoders: one branch processes the entire image \(x_s\), while the other focuses exclusively on the target area \(r_s\).
The generator takes as input the original image and the four wavelets associated with it, the target area of interest, and the output target modality, generating in the output the translated image in the desired domain. This network comprises two branches, the first one for the whole image and the second one for the region of interest. A single branch is composed of an initial encoder, then a set of blocks shared with the other branch, and finally, a decoder to output the desired image. This network stacks convolutional blocks that employ a convolution layer then the instance normalization, and finally a Leaky ReLU activation function. The upsampling blocks instead involve upsampling layers by interpolation suddenly a convolution operation, and, as in the previous block, finally the instance normalization with Leaky ReLU. The discriminator has a similar structure, for each detail regarding the model structure and hyperparameters we refer to the original paper \cite{Chen2021TarGANTG}. Overall, this model comprises $125$ million parameters, being the largest in our experiments. In this case, we introduce the quaternion-valued counterparts as the quaternion TarGAN has not been introduced yet in literature.
\begin{table*}[t]
    \centering
    \caption{Results for task 1 reconstruction on the IXI dataset for real and quaternion-valued models. Average and standard deviation over multiple runs reported. Runs statistically significant with $p < 0.1$.}
    \label{tab:recon}
    \begin{tabular}{lcccc}
        \toprule
        Model & Params & SSIM$\uparrow$ & MSE$\downarrow$ & FID$\downarrow$ \\
        \hline
        CAE \cite{ParcolletICASSP2019a} & 305k & $0.9915 \pm 0.0079$ & $0.0006 \pm 0.0001$ & $0.0046 \pm 0.0035$ \\
        QUAVE-CAE & 305k & $\mathbf{0.9915 \pm 0.0051}$ {(+0\%)} & $\mathbf{0.0006 \pm 0.0001}$ {(+0\%)} & $\mathbf{0.0036 \pm 0.0006}$ {(+21.7\%)} \\
        \hline
        QCAE \cite{ParcolletICASSP2019a}& 78k & $0.9504 \pm 0.0027$ & $0.0015 \pm 0.0001$ & $0.0293 \pm 0.0024$ \\
        QUAVE-QCAE & 78k & $\mathbf{0.9842 \pm 0.0022}$ {(+3.5\%)} & $\mathbf{0.0010 \pm 0.0001}$ {(+33.3\%)} & $\mathbf{0.0126 \pm 0.0034}$ {(+57.1\%)} \\
        \bottomrule
    \end{tabular}
\end{table*}
\begin{table*}[]
    \centering
    \caption{Results for task 2 segmentation on the Kvasir-SEG dataset for real and quaternion-valued models. Average and standard deviation over multiple runs reported. Runs statistically significant with $p < 0.1$.}
    \label{tab:segm}

    \begin{tabular}{lcccc}
        \toprule
        Model & Params & DSC$\uparrow$ & mIoU$\uparrow$ & MAE$\downarrow$ \\
        \hline
        U-Net \cite{Ronneberger2015UNET} & 7.8M & $0.7474 \pm 0.0595$ & $\mathbf{0.6546 \pm 0.0166}$ & $\mathbf{0.0653 \pm 0.0030}$ \\
        QUAVE-U-Net & 7.8M & $\mathbf{0.7726 \pm 0.0025}$ {(+4.1\%)} & $0.6394 \pm 0.0094$ {($-2.3\%$)} & $0.0666 \pm 0.0015$ {(-2\%)} \\
        \hline
        U-Net++ \cite{Zhou2020UNetRS} & 9.2M & $0.7796 \pm 0.0046$ & $0.6379 \pm 0.0147$ & $0.0664 \pm 0.0016$ \\
        QUAVE-U-Net++ & 9.2M & $\mathbf{0.8000 \pm 0.0107}$ {(+2.3\%)} & $\mathbf{0.6729 \pm 0.0109}$ {(+5.4\%)} & $\mathbf{0.0614 \pm 0.0024}$ {(+7.5\%)} \\
        \hline
        QU-Net \cite{ParcolletICASSP2019a} & 1.95M & $0.7668 \pm 0.0116$ & $0.6293 \pm 0.0203$ & $0.0689 \pm 0.0025$ \\
        QUAVE-QU-Net & 1.95M & $\mathbf{0.7728 \pm 0.0102}$ {(+0.8\%)} & $\mathbf{0.6382 \pm 0.0159}$ {(+1.4\%)} & $\mathbf{0.0663 \pm 0.0027}$ {(+3.7\%)} \\
        \hline
        QU-Net++ \cite{ParcolletICASSP2019a} & 2.3M & $0.7857 \pm 0.0078$ & $0.6496 \pm 0.0223$ & $0.0656 \pm 0.3496$ \\
        QUAVE-QU-Net++ & 2.3M & $\mathbf{0.7912 \pm 0.0020}$ {(+0.7\%)} & $\mathbf{0.6526 \pm 0.0039}$ {(+0.4\%)} & $\mathbf{0.0636 \pm 0.0003}$ {(+3.1\%)} \\
        \bottomrule
    \end{tabular}
\end{table*}

\subsection{Metrics}

As objective evaluation metrics in the reconstruction task, we compute the structural similarity index (SSIM), the mean squared error (MSE), and the Fréchet inception distance (FID). While SSIM measures image degradation through structural information, FID estimates how the distributions of real and reconstructed images are far from each other.
Indeed, FID is traditionally used in generative tasks where paired metrics are unavailable. In reconstruction tasks, its ability to capture differences in the distribution of high-level image features can offer complementary insights into reconstruction fidelity, particularly in terms of global structure and perceptual coherence.
By including it alongside SSIM and MSE, we ensure a more comprehensive evaluation of our method ability to reconstruct high-quality images while maintaining both local and global features.
For evaluating the results of the segmentation task instead, we compute the Dice similarity coefficient (DSC), the mean intersection over union (mIoU), and the mean absolute error (MAE).
In order to perform a robust evaluation of the image modality translation task, we employ both a standard FID and an FID proposed in the original work \cite{Chen2021TarGANTG}. 
The standard FID metric measures the distance between the distributions of feature vectors extracted from real and generated images using a pretrained network, commonly the Inception v3 model \cite{szegedy2016rethinking}, which is trained on generic (non-medical) images. 
In the classic implementation, feature vectors from the last pooling layer of the pretrained Inception model are used to calculate the mean and covariance for both the real and generated images.
Instead, the FID implemented by TarGAN \cite{Chen2021TarGANTG} incorporates specific adaptations tailored for the task of translating medical images. To compute the FID, the feature vectors are extracted from the last layer of the encoder of a pre-trained Models Genesis \cite{modelgenesis} network and not the Inception v3 model like in the classic FID. Since Models Genesis \cite{modelgenesis} is based on 3D medical volume data, it is crucial that the input to this network also has a 3D context.
In our image translation task, only 2D translations are performed. To provide the required 3D context for feature extraction, we form 3D volume patches by concatenating every 16 consecutive axial slices, following the receipt in \cite{modelgenesis}. The choice of 16 is based on a balance between capturing sufficient volumetric context and maintaining computational efficiency. Importantly, we ensure that the slices are anatomically consistent, i.e. they are consecutive so that the resulting 3D patch reflects the true spatial continuity of the anatomy. This ensures that the features extracted by Models Genesis are meaningful and appropriate for FID computation.
The average FID score across all modality pairs (e.g., CT $\rightarrow$ T1w, T2w $\rightarrow$ T1w) is reported. {Performance differences may exist between individual translation pairs, for example, translating from CT to T1w typically yields a higher FID score than translating from T1w to CT, owing to inherent contrast differences between these modalities. However, our analysis shows that despite these individual variations, the average FID remains a robust indicator of the overall quality and consistency of the translation process. In this context, the averaged metric effectively captures the general performance of our approach across all modality pairs, demonstrating that our method handles varying degrees of translation difficulty competently. 

\begin{table*}[t]
\centering
\caption{Results for task 3 image modality translation on the CHAOS dataset (image generation metrics) for real and quaternion-valued models. Average and standard deviation over multiple runs reported. Runs statistically significant with $p < 0.1$.}
\begin{tabular}{lccc} 
\toprule
Model  & FID\cite{Chen2021TarGANTG}$\downarrow$ & FID$\downarrow$ & IS$\uparrow$ \\
\midrule
TarGAN \cite{Chen2021TarGANTG}  & $0.0912 \pm 0.0002$ & $0.7009 \pm 0.1181$ & $3.1879 \pm 0.2172$ \\
QUAVE-TarGAN  & $\mathbf{0.0893 \pm 0.0025}$ {(+2.1\%)} & $\mathbf{0.6295 \pm 0.0233}$ {(+10.1\%)} & $\mathbf{3.3452 \pm 0.0185}$ {(+4.9\%)} \\
\midrule
QTarGAN & $\mathbf{0.0881 \pm 0.0021}$ & $0.6513 \pm 0.0144$ & $2.7842 \pm 0.2628$ \\
QUAVE-QTarGAN  & $0.0902 \pm 0.0024$ {($-2.3\%$)} & $\mathbf{0.6253 \pm 0.0860}$ {(+3.9\%)} & $\mathbf{3.0228 \pm 0.0524}$ {(+8.5\%)} \\
\bottomrule
\end{tabular}
\label{tab:targan_results_fid_is}
\end{table*}

\begin{table*}[t]
\centering
\caption{Ablation study for real and quaternion-valued models using different wavelets in QUAVE.}
 \resizebox{\textwidth}{!}{%

\begin{tabular}{c|ccccccccc}
\hline
Data & Model                & Wavelet & SSIM$\uparrow$ & MSE$\downarrow$ & FID\cite{Chen2021TarGANTG}$\downarrow$ & FID$\downarrow$ & DSC$\uparrow$   & mIoU$\uparrow$ & MAE$\downarrow$   \\ 
\hline
\parbox[t]{2mm}{\multirow{4}{*}{\rotatebox[origin=c]{90}{IXI}}} & \multirow{2}{*}{QUAVE-CAE} & DWT &0.991 $\pm$ 0.002 & 0.001 $\pm$ 0.001 & - & 0.010 $\pm$ 0.001 & - & - &  -  \\
       &                  &  QWT    & \textbf{0.991} $\pm$ 0.005 & \textbf{0.001} $\pm$ 0.001 & - & \textbf{0.003} $\pm$ 0.001 & - & - &  -  \\
 &\multirow{2}{*}{QUAVE-QCAE} &DWT& 0.949 $\pm$ 0.015 & 0.001 $\pm$ 0.001 & - & 0.025 $\pm$ 0.003 & - & - &  - \\
 &    &    QWT & \textbf{0.984} $\pm$ 0.002 & \textbf{0.001} $\pm$ 0.001 & - &\textbf{0.012} $\pm$ 0.003 & - & - &  - \\ 
 \cmidrule{1-10}
\parbox[t]{2mm}{\multirow{8}{*}{\rotatebox[origin=c]{90}{KVASIR}}} &\multirow{2}{*}{QUAVE-U-Net} & DWT & - & - &  -  &- & \textbf{0.775}  $\pm$ 0.002 & \textbf{0.643} $\pm$ 0.015 & 0.066 $\pm$ 0.001 \\
 & & QWT & - & - &  - & - &0.772 $\pm$ 0.002 & 0.639 $\pm$ 0.009 & \textbf{0.066} $\pm$ 0.001 \\
&\multirow{2}{*}{QUAVE-QU-Net} & DWT & - & - &  - &- &0.760 $\pm$ 0.015& 0.624 $\pm$ 0.028 & 0.068 $\pm$ 0.004 \\
& & QWT & - & - &  -  & - &\textbf{0.772} $\pm$ 0.010 & \textbf{0.638} $\pm$ 0.015 & \textbf{0.066} $\pm$ 0.002 \\
 \cmidrule{2-10}

&\multirow{2}{*}{QUAVE-U-Net++}   & DWT & - & - &  - & - &0.781 $\pm$ 0.011 & 0.649 $\pm$ 0.016 & 0.064 $\pm$ 0.003 \\
 & & QWT & - & - &  - &- &\textbf{0.798} $\pm$ 0.010& \textbf{0.672} $\pm$ 0.010 & \textbf{0.061} $\pm$ 0.002 \\
&\multirow{2}{*}{QUAVE-QU-Net++}&DWT& - & - &  -  &- & 0.782 $\pm$ 0.019& 0.645 $\pm$ 0.026 & 0.065 $\pm$ 0.004 \\
 & &QWT & - & - &  - &- &\textbf{0.791} $\pm$ 0.001 &  \textbf{0.652} $\pm$ 0.003& \textbf{0.063} $\pm$ 0.002 \\
\cmidrule{1-10}

 \parbox[t]{2mm}{\multirow{4}{*}{\rotatebox[origin=c]{90}{CHAOS}}} &\multirow{2}{*}{QUAVE-TarGAN} & DWT  & - & - & 0.092 $\pm$ 0.002 & 0.684 $\pm$ 0.040 &  0.888 $\pm$ 0.201 & 0.802 $\pm$ 0.031& 0.048 $\pm$ 0.001 \\
 & &QWT & - & - & \textbf{0.089} $\pm$ 0.002 & \textbf{0.629} $\pm$ 0.023  &  \textbf{0.905} $\pm$ 0.436 & \textbf{0.832} $\pm$ 0.067 & \textbf{0.048} $\pm$ 0.001 \\

&\multirow{2}{*}{QUAVE-QTarGAN} & DWT & - & - & \textbf{0.087} $\pm$ 0.001 & 0.649 $\pm$ 0.069 & \textbf{0.836} $\pm$  0.012 & 0.724 $\pm$ 0.017 & 0.050 $\pm$ 0.001\\
& & QWT & - & - & 0.090 $\pm$ 0.002 & \textbf{0.625} $\pm$ 0.086  & 0.830 $\pm$ 0.076 & \textbf{0.736} $\pm$ 0.009 & \textbf{0.048} $\pm$ 0.001  \\ 
\bottomrule
\end{tabular}
}
\label{tab:recon_ablation}

\end{table*}

\section{Results and Discussion}
\label{sec:exp}
In this section, we present the selected quaternion sub-bands we employ in our tests and the empirical evaluation we accomplish for validating our theoretical claims. We perform multiple runs with different seeds for each experiment and we report in the tables the average score and the standard deviation for each metric. We initialized the neural networks weights (both the 2D convolutional and the linear layers) using the well-known Kaiming and He initialization with a uniform distribution and the biases to zero.

\subsection{Evaluation}
\subsubsection{Reconstruction on IXI dataset}
Table~\ref{tab:recon} shows the average and the standard deviation over different runs of the objective metrics results for the IXI dataset.
The models that employ the proposed preprocessing (QUAVE) enhance the scores with both real and quaternion-valued models.
These results prove that QUAVE can boost model learning without requiring modification or manipulation of models or any kind of data augmentation. Indeed, features extracted with the QWT capture salient characteristics of the image that are crucial for proper reconstruction.

\subsubsection{Segmentation on Kvasir-SEG dataset}
Table~\ref{tab:segm} reports instead objective metrics results for the segmentation task on the Kvasir-SEG dataset with RGB images. It is worth noting that the Kvasir-SEG dataset comprises RGB images, thus the real-valued U-Net is trained with three-channel color samples. Therefore, we prove that our technique, QUAVE, increases model performance also with respect to color images. Hence, QUAVE can be generalized to grayscale and color images as well, improving the accuracy of involved networks.

\subsubsection{Image modality translation on CHAOS dataset}
Table~\ref{tab:targan_results_fid_is} and Table~\ref{tab:targan_results_dsc_s_miou_mae} report objective metrics results for the image modality translation task on the CHAOS dataset with both image generation and segmentation metrics. 
The segmentation results presented in Table~\ref{tab:targan_results_dsc_s_miou_mae} provide insight into the quality of structural preservation in the translated images. This aspect is inherently linked to the TarGAN architecture and methodology.

As in previous experiments, we apply the proposed QUAVE to data for extraction, to both the original TarGAN\cite{Chen2021TarGANTG} and a quaternion version referred to as QTargan. 
To further stabilize GAN training, we also added to the original network the Spectral Normalization to limit the discriminator sensitivity and ensure a more stable training process \cite{miyato2018spectral}.
We can observe that both the DWT and QUAVE improve the performance with respect to the original network. The FID\cite{Chen2021TarGANTG} proposed in the original TarGAN work tends to reward the images that best translate meaningful features in medical images, and our approach QUAVE beats the original processing model. Even if we consider the classical FID, QUAVE performs best among the original work. We also show this effect on a visual comparison among the predicted translation in Fig.~\ref{fig:targan_results} on randomly chosen samples. We display the superiority of the translation of QUAVE using red boxes. The boxes help to indicate the artifacts in the translation committed by the other methods. To conclude, we show that the data-agnostic QUAVE can be easily generalized to multimodality datasets, improving the performance of large-scale models too.

\subsection{Ablation study}
We adopt the QWT for our framework, consequently, we show an ablation study instead of using the DWT in QUAVE. This is possible with only a slight change in the output size of the final layers of QUAVE. 
From the scores in Table~\ref{tab:recon_ablation}, we can observe that both the DWT and QUAVE approaches have a high impact on model performance, improving scores with respect to baselines. However, the best scores are, again, achieved by the models with the QUAVE method with QWT wavelet, proving its effectiveness in most of all the different tasks.

\section{Conclusion}
\label{sec:con}
In this paper, we have introduced QUAVE a generalizable, shift-invariant, and data-agnostic framework, for exploiting salient features in medical images. The method is based on the quaternion wavelet transform and on a novel technique for extracting informative sub-bands within the Quaternion Wavelet Transform for neural models which improves networks' performance in a variety of tasks, while improving quaternion and hypercomplex models' generalizability. On a thorough experimental evaluation with different models in three distinct tasks, we have demonstrated the effectiveness and the generalization ability of our approach, which is able to improve performance according to a variety of metrics in each task we consider. Moreover, we have shown the flexibility of the proposed methods that can be involved in every medical and non-medical dataset, whether with grayscale, RGB images, or comprising different modalities.
As future work, QUAVE may be defined to directly operate with 3D volumes of medical imaging inputs and outputs, so as to enable 3D downstream models in both real- and quaternion-valued domains.

\bibliographystyle{ieeetr}
\bibliography{arxiv_biblio}

\end{document}